\documentclass{article}
\def\DR{\rm I\kern-1.45pt\rm R}
\def\DC{\kern2pt {\hbox{\sqi I}}\kern-4.2pt\rm C}

\newcommand{\ba}{\begin{array}}
\newcommand{\ea}{\end{array}}
\newcommand{\be}{\begin{equation}}
\newcommand{\ee}{\end{equation}}
\newcommand{\bea}{\begin{eqnarray}}
\newcommand{\eea}{\end{eqnarray}}
\newcommand{\bi}{\begin{itemize}}
\newcommand{\ei}{\end{itemize}}

\usepackage{amscd,amsmath,amssymb}

\begin{document}\begin{center}
{\bf \Large   Two--center quantum MICZ--Kepler system and the Zeeman effect in the charge-dyon system}\\
\vspace{0.5 cm} {\large Stefano Bellucci$^1$ and Vadim
Ohanyan$^2$}
\end{center}
\noindent
$\;^1${\it INFN-Laboratori Nazionali di Frascati, Via E. Fermi 40, 00044 Frascati, Italy}\\
$\;^2$ {\it Yerevan State University, A.Manoogian, 1, Yerevan,
375025 Armenia\\
 Yerevan Physics Institute, Alikhanian Br.2,
Yerevan, 375036, Armenia}\\
{\sl E-mails: bellucci@lnf.infn.it,
  ohanyan@yerphi.am}
\begin{abstract}
The quantum two-center MICZ--Kepler system is considered in the
limit when one of the interaction centers is situated at infinity,
which leads to homogeneous electric and magnetic fields appearing
in the system. The emerging system admits separation of variables
in the Schr\"{o}dinger equation and is integrable at the classical
level.
%In the physical context this system describes the
%charge--dyon system subjected to homogeneous electric and magnetic
%fields parallel to each other. Another important feature which
%guarantees the separation of variables is the additional potential
%terms, oscillator like and proportional to $\cos \theta$.
The
first order corrections to the unperturbed spectrum of the
ordinary MICZ--Kepler system are calculated. Particularly, the
linear Zeeman--effect and effects of MICZ-terms are analyzed. The
possible realizations of the system in some quantum dots are
considered.
\end{abstract}

\section{Introduction}
The MICZ--Kepler system describes an electrically charged scalar
particle moving in the field of static Dirac dyon(s), i.e. a
particle carrying both electrical and magnetic charges. This system
was originally proposed by Zwanziger \cite{Z} and McIntosh and
Cisneros independently \cite{MIC}. In their papers the corresponding
system was obtained from the ordinary Coulomb problem by replacing
the interaction center by a dyon which has, beside the scalar
potential, the vector-potential as well, and by introducing into the
Hamiltonian the additional centrifugal-like term, $s^2/2\mu r^2$,
where $s=eg$ is the monopole number (we use $c=\hbar=1$ units), $e$
is the electric charge of the probe particle, $g$ is the magnetic
charge of the static dyon and $\mu$ is the mass of the probe
particle. According to the Dirac quantization rule the monopole
number can take integer or half--integer values: $s=,0,\pm 1/2, \pm
1...$ The corresponding system is described by the following
Hamiltonian: \bea {\mathcal{H}}_{MICZ}=\frac{1}{2
\mu}\left({\mathbf{p}}-e{\mathbf{A}}_g \right)^2+\frac{s^2}{2 \mu
r^2 }+\frac{e q}{r}, \label{H0}  \eea where\bea \mbox{rot}
{\mathbf{A}}_g=\frac  {g{\mathbf{r}}}{r^3}. \eea
 Due to this
additional potential term the emerging system inherits all
distinguished features of the underlying Coulomb system. For
instance, the MICZ-Kepler system, beside the general rotational
symmetry generated by the algebra of angular momentum components
($so(3)$), has also a higher $so(4)$-hidden symmetry which is
connected to the conservation of the Laplace--Runge--Lentz vector.
The expressions for these conserved quantities are very similar to
those of the Coulomb problem
\begin{eqnarray}
{\mathbf{J}}={\mathbf{r}} \times \left({\mathbf{p}}-e
{\mathbf{A}}\right) +s\frac{{\mathbf{r}}}{r}, \quad
{\mathbf{I}}=\frac{1}{\mu}\left({\mathbf{p}}-e
{\mathbf{A}}\right)\times {\mathbf{J}}-eq\frac{{\mathbf{r}}}{r}.
\label{JI}
\end{eqnarray}
As one can see, only the additional spin-like term appears in the
expression for the orbital momentum. The hidden symmetry of the
MICZ--Kepler system leads to the possibility of separation of
variables in several coordinate systems, e. g. in spherical and
parabolic ones \cite{mic_par}. The origin of the additional
potential term (MICZ--term) could be interpreted as the
interaction energy between the spherically symmetric magnetic
field of Dirac monopole and the magnetic moment of the probe
particle, ${\mathbb{\mathcal{M}}}=\frac{e}{2 \mu c}{\mathbf{J}}$,
where ${\mathbf{J}}$ is given by Eq. (\ref{JI}). However,
additional part of the angular momentum vector proportional to the
monopole number $s$ should be rather assigned to the monopole
electromagnetic field by itself but not to the particle. Thus,
such a clear and natural, at the first glance, explanation of the
origin of MICZ-term is not quite correct. On the other hand, this
term can be obtained both on classical and quantum level when one
reduce the four--dimensional isotropic oscillator with the aid of
so--called Kustaanheimo--Stiefel transformations
\cite{mla}.

 The shape of the classical trajectories of the MICZ--Kepler
 system coincides with these of the ordinary Coulomb one, however
 in contrast to the latter the orbital plane are non orthogonal to
 the angular momentum vector. The angle between ${\mathbf{J}}$ and
 orbital plane normal satisfy the relation:
 \bea
 \cos \chi=\frac{s}{|{\mathbf{J}}|},
 \eea
 which immediately followed from Eq. (\ref{JI}). The properties of
 the solutions of the quantum mechanical MICZ--Kepler problem
 undergo minor modifications with respect to those of the
 underlying Coulomb one. Namely, it has the same spectrum, given
 in terms of the same quantum numbers. The only difference is in
 the range of the possible values of angular momentum. The
 presence of the monopole with monopole number $s$ shifts upward
 the allowed values of orbital and magnetic quantum numbers $j$
 and $m$: $j=|s|, |s|+1, |s|+2,...$ There is also more general
 statement concerning the incorporation of Dirac monopoles in
 spherically symmetric mechanical systems. Any spherically
 symmetric system (without monopoles) defined on the conformally
 flat space with metric $G(r)$ will preserve its spectral
 properties except the possible range of angular momentum values
 mentioned above, if adding the monopole potential will accompany
 with the following modification of potential term \cite{Lec, MNY,
 NO_TMP, bny1}:
 \bea
 U(r)\rightarrow U(r)+\frac{s^2}{2 G(r) \mu r^2}
 \eea
 Another difference between Coulomb systems and their
 MICZ--counterparts manifests itself in the behavior in external
 fields. So, one can observe the modification of the selection
 rules in the dipole transitions \cite{dipol,dipole}. The
 investigations of the generalizations of the MICZ--Kepler system
 to the higher dimensions and/or curved spaces as well revealed their
 close similarity to the underlying Coulomb systems
 \cite{NO_TMP,MIC1,MIC_hyp,MIC_any}. Multi--center generalization
 of MICZ--Kepler systems has been found recently in Ref.
 \cite{NO_TMP,KNO07}. The Hamiltonian which describes the charged
 particle moving in the electromagnetic field of $n$ Dirac dyons
 fixed in the Euclidian space points with radius-vectors
 ${\mathbf{a}}_i$, $i=1,...,n$ reads:
 \be \label{H00}
\mathcal{H}=\frac{1}{2 \mu} \left(\mathbf{p}-e \sum_{i=1}^n
\mathbf{A}_{g_i}(\mathbf{r}-\mathbf{a}_i)
  \right)^2 + e\sum_{i=1}^n
  \frac{q_i}{|\mathbf{r}-\mathbf{a}_i|}+\frac{e^2}{2
  m}\left(\sum_{i=1}^n\frac{g_i}{|\mathbf{r}-\mathbf{a}_i|}
  \right)^2, \quad
\mbox{rot}\mathbf{A}_{g_i}(\mathbf{r})=\frac{g_i \mathbf{r}}{r^3}
\ee where $q_i(g_i)$ are the electric(magnetic) charges of the the
$i$-th dyon. The last term in the expression generalizes the
MICZ-term for the multi-center situation. In full analogy with the
corresponding pure Coulomb system two--center case is integrable on
the classical level \cite{NO_TMP,KNO07}, i .e. it allows separation
of variables in the Hamilton--Jacobi equation in elliptic
coordinates. The limiting case of the two--center MICZ--Kepler
system when one of the dyons is situated at infinity which results
in the homogeneous electric and magnetic fields, is also integrable
on classical level. The MICZ--term corresponding to this dyon
results in its turn in the quadratic potential. Corresponding
Hamilton--Jacobi equation become separable in parabolic coordinates.
Because of presence of homogeneous electric and magnetic fields
imposed to the one--center MICZ--Kepler (charge--dyon) system one
can call it MICZ--Kepler--Zeeman--Stark system. In this paper we
consider some quantum mechanical issues of the
MICZ--Kepler--Zeeman--Stark system.In addition to the pure stark
effect in charge--dyon system considered earlier in
Refs.\cite{MNP_04,dipole}, we present perturbative corrections to
the spectrum for the charge--dyon system corresponding to the
homogeneous magnetic field(linear Zeeman effect).

 It is also noteworthy that multi--center MICZ--Kepler system
 allows ${\mathcal{N}}=4$ supersymmetric extension when magnetic
 and electric charges of all dyons satisfy the trivial
 Dirac--Schwinger--Zwanziger condition\cite{KNO07}
 \be g_i q_j-g_j q_i=0.
 \ee
 The corresponding  ${\mathcal{N}}=4$ supersymmetric mechanical
 systems were constructed in Refs. \cite{iva,bk} and with a little
 different approach leading to the MICZ--Kepler system on the
 three--dimensional sphere in Ref. \cite{BKO07}.\footnote{For a
 unified superfield formulation of N=4 off-shell supermultiplets
 in one space-time dimension see \cite{new}. For earlier work see \cite{old}.}

  The paper is organized as follows. In the second section we
  derive the Schr\"{o}dinger equation for the
  MICZ--Kepler--Zeeman--Stark system in parabolic coordinates. The
  solution of the unperturbed problem is presented and calculated
  the first and second order corrections in $B$ and
  ${\mathcal{E}}$ The average dipole and magnetic moments acquired the system are also calculated.
  In the third section the related quantum dot
  models are discussed. Some final remarks are presented in the
  conclusion. The appendix contains some technical points.
\section{Two-center MICZ--Kepler system and quantum MICZ--Kepler--Stark--Zeeman System}
The Hamiltonian of the two center MICZ--Kepler system on Euclidean
space has the form \cite{KNO07}
\begin{eqnarray}
{\mathcal{H}}=\frac{1}{2}\left({\mathbf{p}}-e
{\mathbf{A}}_{g_1}({\mathbf{r}}-{\mathbf{r_1}})-e{\mathbf{A}}_{g_2}({\mathbf{r}}-{\mathbf{r_2}})\right)^2+\frac{1}{2}\left(
\frac{s_1}{|{\mathbf{r}}-{\mathbf{r_1}}|^2}+\frac{s_2}{|{\mathbf{r}}-{\mathbf{r_2}}|^2}\right)^2+e
\left(
\frac{q_1}{|{\mathbf{r}}-{\mathbf{r_1}}|}+\frac{q_2}{|{\mathbf{r}}-{\mathbf{r_2}}|}\right),\label{2.1}
\end{eqnarray}
where
\begin{eqnarray}
{\mathbf{A}}_g({\mathbf{r}})=g\frac{{\mathbf{n}}\times{\mathbf{r}}}{r
\left( r-{\mathbf{n}}{\mathbf{r}} \right)}\label{2.2}
\end{eqnarray}
is the vector--potential of the Dirac dyon with magnetic charge
$g$, ${\mathbf{n}}$ is the unit vector pointing to the singularity
line. Choosing an appropriate gauge for the monopole
vector-potential (\ref{2.2}),
\begin{eqnarray}
A_r=A_{\theta}=0, \quad A_{\varphi}=g \cos \theta, \label{2.3}
\end{eqnarray}
and supposing that the dyons be fixed on the $z$-axis at points
$(0,0,a)$ and $(0,0,-a)$ respectively, one arrives at the
following expression for the Hamiltonian (\ref{2.1}) in spherical
coordinates:
\begin{eqnarray}
{\mathcal{H}}=\frac{1}{2}\left(p_r^2+\frac{p_{\theta}^2}{r^2}
+\frac{\left(p_{\varphi}-s_1 \cos \theta_1 -s_2 \cos \theta_2
\right)^2}{r^2 \sin^2
\theta}\right)+\frac{1}{2}\left(\frac{s_1}{r_1}+\frac{s_2}{r_2}\right)^2+\frac{eq_1}{r_1}+\frac{eq_2}{r_2},
\label{2.4}
\end{eqnarray}
where $r_{1,2}=\sqrt{x^2+y^2+(z \pm a)^2}$.
 As it was shown in Ref. \cite{KNO07} this system on the classical
level admits separation of variables in elliptic coordinates, what
leads to the integrable system of Hamilton--Jacobi equations. The
two--center MICZ--Kepler system has an important limiting case,
when one of the interaction centers is situated at infinity. The
field of such a dyon results in the homogeneous electric and
magnetic fields being parallel to each other. If one takes for the
vector potential of the homogeneous magnetic field the following
gauge:
\begin{eqnarray}
A_r=A_{\theta}=0, \quad A_{\varphi}=\frac{1}{2}B r^2 \sin^2
\theta, \label{2.5}
\end{eqnarray}
 then the corresponding Hamiltonian in spherical coordinates reads
\begin{eqnarray}
{\mathcal{H}}=\frac{1}{2}\left(p_r^2+\frac{p_{\theta}^2}{r^2}
+\frac{\left(p_{\varphi}-s \cos \theta -\frac{1}{2}e B r^2 \sin^2
\theta \right)^2}{r^2 \sin^2
\theta}\right)+\frac{1}{2}\left(\frac{s}{r}+e B
z\right)^2+\frac{eq}{r}-e {\mathcal{E}} z, \label{2.6}
\end{eqnarray}
where $B$ and ${\mathcal{E}}$ stand for the modulus of the
magnetic and electric field respectively which are pointed in the
$z$ direction. The classical case is characterized by the
separation of variables in the Hamilton--Jacobi equation in
parabolic coordinates given by the following relations:
\begin{eqnarray}
\xi=r+z, \quad \eta=r-z, \quad \varphi=\arctan \frac{y}{x}.
\label{2.7}
\end{eqnarray}
The same feature holds in the quantum case for the Schr\"{o}dinger
equation for the Hamiltonian (\ref{2.6}) as well. The quantum
Hamiltonian resulting from Eq. (\ref{2.6}) reads
\begin{eqnarray}
{\mathcal{H}}=-\frac{1}{2}\triangle+i \left( \frac{s \cos
\theta}{r^2 \sin^2 \theta}+ \omega_B
\right)\frac{\partial}{\partial \varphi}+\frac{s^2}{2 r^2 \sin^2
\theta}+\frac{\omega_B^2}{2}\left(\rho^2+4 z^2
\right)+3s\omega_B\cos \theta+\frac{e q}{r}-e {\mathcal{E}}z,
\label{28}
\end{eqnarray}
where $\omega_B=\frac{e B}{2}$ is the cyclotron frequency in our
units ($c=\hbar=\mu=1$) and $\rho^2=r^2\sin^2 \theta=x^2+y^2$.
Being rewritten in the parabolic coordinates given by Eqs.
(\ref{2.7}), the corresponding Schr\"{o}dinger equation takes the
form
\begin{eqnarray}
&&\frac{4}{\xi+\eta}\left( \frac{\partial}{\partial \xi}\left( \xi
\frac{\partial \Psi}{\partial \xi} \right)+
\frac{\partial}{\partial \eta}\left( \eta \frac{\partial
\Psi}{\partial \eta} \right) \right)+\frac{1}{\xi \eta} \left(
\frac{\partial^2}{\partial \varphi^2}-s^2\right)\Psi+2 i
\left(\frac{s}{\xi+\eta}\left( \frac{1}{\xi}-\frac{1}{\eta}\right)
-\omega_B\right)\frac{\partial\Psi}{\partial\varphi}+ \\ \nonumber
&&\left( 2 E-\frac{\omega_B^2\left(\xi^3+\eta^3
\right)}{\xi+\eta}-6s\omega_B \frac{\xi-\eta}{\xi+\eta}-\frac{4 e
q}{\xi+\eta}+ e
{\mathcal{E}}\frac{\xi^2-\eta^2}{\xi+\eta}\right)\Psi=0.\label{2.9}
\end{eqnarray}
Hereafter in order to deal with discrete spectrum we suppose the
electric charge $e$ of the probe particle to be negative. Seeking
for the eigenfunctions in the form
\begin{eqnarray}
\Psi=f_1(\xi)f_2(\eta)e^{i m \varphi}, \label{2.10}
\end{eqnarray}
where $m$ is the magnetic quantum number, and dividing the
equation on $\frac{4f_1f_2}{\xi+\eta}$, one obtains
\begin{eqnarray}
 \frac{d}{d \xi}\left( \xi \frac{d f_1}{d
\xi} \right)\frac{1}{f_1}+ \frac{d}{d \eta}\left( \eta \frac{d
f_2}{d \eta} \right)\frac{1}{f_2}+U(\xi)+V(\eta)=-|e| q
\end{eqnarray}
where
\begin{eqnarray}
&&U(\xi)=-\frac{s_+^2}{4 \xi}-\frac{\omega_B^2}{4}\xi^3+\frac{e
{\mathcal{E}}}{4}\xi^2+W_-\xi, \\ \nonumber
&&V(\eta)=-\frac{s_-^2}{4 \eta}-\frac{\omega_B^2}{4}\eta^3-\frac{e
{\mathcal{E}}}{4}\eta^2+W_+\eta
\end{eqnarray}
and the following notations are introduced $s_{\pm}=m \pm s$,
$W_{\pm}=1/4 (2E+2m \omega_B \pm 6s\omega_B)$. Separating the
variables $\xi$ and $\eta$ one obtains the system of equation for
the functions $f_1$ and $f_2$
\begin{eqnarray}
\frac{d}{d \xi}\left( \xi \frac{d f_1}{d \xi} \right)+\left(|e| q
\beta_1-\frac{s_+^2}{4 \xi}-\frac{\omega_B^2}{4}\xi^3+\frac{|e|
{\mathcal{E}}}{4}\xi^2+W_-\xi\right)f_1=0, \\ \nonumber \frac{d}{d
\eta}\left( \eta \frac{d f_2}{d \eta} \right)+\left(|e| q
\beta_2-\frac{s_-^2}{4 \eta}-\frac{\omega_B^2}{4}\eta^3-\frac{|e|
{\mathcal{E}}}{4}\eta^2+W_+\eta \right)f_2=0,\label{SE_full}
\end{eqnarray}
where the separation constants $\beta_1$ and $\beta_2$ satisfy the
condition
\begin{eqnarray}
\beta_1+\beta_2=1\label{beta}
\end{eqnarray}
\subsection{One-center MICZ--Kepler system in parabolic coordinates}
Note that the case of the one-center MICZ-Kepler ($B=0,
{\mathcal{E}}=0$) system completely coincides with the
quantum-mechanical problem of the hydrogen atom. \footnote{
Related supersymmetric systems with Dirac monopoles as well as the
 details of the Hamiltonian reduction technique have been considered in
\cite{BKS}.} It
leads to the same spectrum. The only difference consists in the
allowed values of magnetic quantum number, which are just shifted
by s. Now one can regard the terms corresponding to the
homogeneous electric and magnetic fields as perturbations with
respect to the integrable case of the one center MICZ--Kepler
Hamiltonian which leads to the following equations\cite{mic_par}:
\begin{eqnarray}
\frac{d}{d \xi}\left( \xi \frac{d f_1}{d \xi}
\right)+\left(\frac{1}{2} E \xi-\frac{s_+^2}{4 \xi}+|e| q \beta_1 \right)f_1=0, \\
\nonumber \frac{d}{d \eta}\left( \eta \frac{d f_2}{d \eta}
\right)+\left(\frac{1}{2} E\eta-\frac{s_-^2}{4\eta}+|e| q \beta_2,
\label{SE_C} \right)f_2=0
\end{eqnarray}
which are almost identical to those for the hydrogen atom in
parabolic coordinates (see for example \cite{QM}). We consider
only the discrete spectrum of the unperturbed problem ($E<0$). In
this case it is convenient to pass to the new variables
$n=1/\sqrt{-2E}$, $\rho_1=\xi/n$ and $\rho_2=\eta/n$. Let us write
down the equation for the function $f_1$ in these coordinates
\begin{eqnarray}
\frac{d^2 f_1}{d\rho_1^2}+\frac{1}{\rho_1}\frac{d f_1}{d
\rho_1}+\left(
-\frac{1}{4}+\frac{1}{\rho_1}\left(n_1+\frac{1}{2}\left(
|s_+|+1\right)\right)-\frac{s_+^2}{4
\rho_1^2}\right)f_1=0,\label{eq_f1}
\end{eqnarray}
where we introduce the notations and put for simplicity $|e|q=1$
(natural units)
\begin{eqnarray}
n_1=n  \beta_1-\frac{1}{2}\left(|s_+|+1\right), \quad n_2=n
\beta_2-\frac{1}{2}\left(|s_-|+1\right). \label{n1n2}
\end{eqnarray}
The corresponding solution for $f_2$ can be obtained by replacing
$\beta_1$ by $\beta_2$ and $s_+$ by $s_-$. Taking into account the
long- and short-distance asymptotic behavior of the solutions one
can seek the function $f_1$ in the following form: \cite{QM}
\begin{eqnarray}
f_1(\rho_1)=e^{-\rho_1 /2} \rho_1^{|s_+|/2}F_1(\rho_1),
\label{f-F}
\end{eqnarray}
which leads to the equation for the confluent hypergeometric
function for $F_1(\rho_1)$
\begin{eqnarray}
\rho_1 \frac{d^2F_1}{d \rho_1^2}+\left(|s_+|+1-\rho_1
\right)\frac{d F_1}{d \rho_1}+n_1 F_1=0. \label{eq_F1}
\end{eqnarray}
Thus, the functions $F_1$ and $F_2$ which satisfy the finiteness
condition read
\begin{eqnarray}
F_1(\rho_1)=F\left( -n_1, |s_+|+1; \rho_1 \right), \quad
F_2(\rho_1)=F\left( -n_2, |s_-|+1; \rho_2 \right), \label{F1F2}
\end{eqnarray}
where $n_1$ and $n_2$ must be non-negative integers. So, from Eq.
(\ref{n1n2}) one obtains that the principal quantum number for the
one-center MICZ-Kepler system is
\begin{eqnarray}
n=n_1+n_2+\frac{|m+s|+|m-s|}{2}+1, \label{n}
\end{eqnarray}
Thus, for the unperturbed part of the Hamiltonian (\ref{2.6}),
each stationary state of the discrete spectrum in parabolic
coordinates is characterized by three integer quantum numbers,
i.e. the parabolic quantum numbers $n_1$ and $n_2$ and the
magnetic quantum number $m$. The normalized eigenfunctions are
\cite{QM}
\begin{eqnarray}
\Psi_{n_1,n_2,m}(\xi,
\eta,\varphi)=\frac{1}{n^2\sqrt{\pi}}f_{n_1,m+s}\left(\xi/n
\right)f_{n_2,m-s}\left(\xi/n \right)e^{i m \varphi},\label{PSI}
\end{eqnarray}
where
\begin{eqnarray}
f_{k,p}\left( \rho \right)=\frac{1}{|p|!}\sqrt{\frac{(k+|p|)!}{k
!}}e^{-\rho/2}\rho^{|p|/2}F\left(-k,|p|+1; \rho \right).
\label{fkp}
\end{eqnarray}
\subsection{The perturbative first order corrections and linear Zeeman splitting}
One can regard all terms, corresponding to magnetic and electric
field as small perturbations. Thus, we have the unperturbed
Hamiltonian of the one-center MICZ--Kepler system with exactly
known stationary state eigenfunctions (\ref{PSI}) and the
perturbation given by the operator
\begin{eqnarray}
{\mathcal{W}}=|e|{\mathcal{E}}z-\frac{3 s |e| B}{2} \cos \theta
+\frac{ e^2B^2}{8}\left(\rho^2+4z^2
\right)=\frac{|e|{\mathcal{E}}}{2}\left(\xi-\eta \right)-\frac{3 s
|e| B}{2} \frac{\xi-\eta}{\xi+\eta}+\frac{
e^2B^2}{8}\frac{\xi^3+\eta^3}{\xi+\eta}. \label{pert}
\end{eqnarray}
There is also a constant shift in the energy level, caused by the
magnetic field and proportional to the magnetic quantum number
$m$, i.e. $-m\omega_B$. So, the first order correction to the
stationary unperturbed hydrogen--like spectrum\cite{mic_par}
\begin{eqnarray}
E_{n_1n_2m}^{(0)}=-\frac{1}{2
\left(n_1+n_2+1+\frac{|m+s|+|m-s|}{2} \right)^2},
\end{eqnarray}
reads
\begin{eqnarray}
\triangle E_{n_1n_2m}^{(1)}= \langle
n_1,n_2,m|{\mathcal{W}}|n_1,n_2,m \rangle,
\end{eqnarray}
Thus
\begin{eqnarray}
E_{n_1n_2m}^{(1)}=E_{n_1n_2m}^{(0)}-m\omega_B+\triangle
E_{n_1n_2m}^{(1)},
\end{eqnarray}
where
\begin{eqnarray}
\triangle E_{n_1n_2m}^{(1)}=\frac{|e|{\mathcal{E}}}{4}{\mathcal
I}_2^--\frac{3 s|e| B }{4 n}{\mathcal I}_1^-+\frac{n e^2
B^2}{8}{\mathcal I}_3^+ \label{III}
\end{eqnarray}
and ${\mathcal I}_k^{\pm}$ stand for the integrals of
hypergeometric functions presented in Appendix. Here we restrict
ourselves with the first-order  corrections in  ${\mathcal{E}}$
and $B$ (we will suppose ${\mathcal{E}}$ and $B$ to be of the same
order). Thus, \bea
E^{(1)}_{n_1n_2m}=-\frac{1}{2n^2}+\frac{3}{2}|e|{\mathcal{E}}(n
n_--\frac{ms}{3})-\frac{1}{2}|e|B (\frac{3 s n_-}{n}-m)+O(B^2),
\label{E1} \eea where \bea n_-=\frac{|m+s|-|m-s|}{2}+n_1-n_2.
\label{n-}\eea Putting $B=0$ one obtains the well known result of
purely Stark--effect in the charge--dyon system \cite{dipole,
MNP_04}. The third term in the expression corresponds to the
linear Zeeman--effect. As usual, in this approximation  magnetic
field removes the degeneracy with respect to the magnetic quantum
number $m$ and induce the magnetic moment in the system. The
average magnitude of induced magnetic moment could be obtained by
taking the derivative of Eq. (\ref{E1})with respect to $B$.
However, the term $\frac{3 s |e| B}{2} \cos \theta$ in the
operator (\ref{pert}) actually comprise of two parts of different
origin. The first is $ \frac{s|e|B}{2} \cos \theta$ which is
originated from the magnetic field in the kinetic term. The second
part is $ s|e|B \cos \theta$ coming from the additional potential
term (MICZ--term). Indeed, the MICZ--term in this system just has
the coefficient which is coincide with magnetic field by magnitude
but physically is not identical with it. Thus, in order to
calculate the induced magnetic moment of the charge--dyon system
in the external magnetic field one must subtract from
$E^{(1)}_{n_1n_2m}$ the impact of the MICZ--term prior to taking
the derivative:

 \bea
\overline{\mathcal{M}}_z&=&-\frac{\partial \left(
E^{(1)}_{n_1n_2m}-s|e|B\langle n_1,n_2,m| \cos \theta|n_1,n_2,m
\rangle\right)} {\partial B}=\frac{1}{2}|e| (\frac{ s n_-}{n}-m)
\\ \nonumber &=&\mu_B\left(s
\frac{|m+s|-|m-s|+2(n_1-n_2)}{|m+s|+|m-s|+2(n_1+n_2+1)}-m \right),
\label{Mz}\eea where $\mu_B=\frac{1}{2}|e|$ is the Bohr magneton
in our units. Taking into account above mentioned properties of
the linear in $B$ corrections one can write down the spectrum in
the following form: \bea
E_{n_1n_2m}^{(1)}=E_{n_1n_2m}^{(0)}-\overline{d}_z{\mathcal{E}}
-\overline{\mathcal{M}}_zB+\frac{s|e| n_- B}{n}. \label{sp2} \eea
The last term originated from the MICZ-term and \bea
\overline{d_z}=-\frac{\partial \triangle
E_{n_1n_2m}^{(1)}}{\partial
{\mathcal{E}}}=-\frac{3}{2}|e|{\mathcal{E}}(n n_--\frac{ms}{3}),
\label{dip} \eea is the average dipole momentum in linear
approximation, calculated in Ref. \cite{MNP_04}
 For the
ground state of unperturbed system (charge-dyon system) which is
characterized by $n_1=n_2=0$, $|m|<|s|$ $\Rightarrow$ $n=|s|+1$,
$n_-=m  {\rm sgn} (s)$, $m= -|s|, -|s|+1,...,|s|-1, |s|$ one
obtains \bea \overline{{\mathcal{M}}_z}=-\mu_B \frac{ m}{1+|s|}.
\label{Mz_gr}\eea
 Thus, the linear Zeeman--effect in the charge--dyon system
 removes the $(2|s|+1)$-fold degeneracy by $m$ of the ground
 state. The induced magnetic momentum for positive values of m is
 always negative and goes to zero at $|s| \to \infty$. At $s=0$
 one get the expected value $\overline{\mathcal{M}}_z=-\mu_b m$
 corresponding to the ordinary Zeeman splitting in the hydrogen
 atom.
 Hence, for the ground state of the system under consideration one
 can write
 \bea
 E_0^{(0)}=-\frac{1}{2(1+|s|)^2}+m|e| \mbox{sgn} s \left(
 |s|+\frac{3}{2}\right){\mathcal{E}}+\frac{1}{2}|e|\frac{m
 }{1+|s|}B+|e|\frac{m |s|}{1+|s|} B.
 \eea
 As one can see, the sign of the linear Zeeman effect depends only
 on the sign of the magnetic quantum number $m$, whereas the
 linear Stark effect depends on the relative sign of $m$ and $s$.
If we suggest that coefficient in the oscillatory potential
$\frac{\omega_0^2}{2}(\rho^2+4z^2)$ in Eq. (\ref{pert}) is
independent of $B$ and is of the same order with ${\mathcal{E}}$ and
$B$ then according to the formula (\ref{III}) we will get the
corresponding first order correction to the ground state energy in
the following form (see Appendix) \bea \triangle
\tilde{E}_0^{1}=\frac{\omega_0^2}{2}(1+|s|)((m^2+s^2)(|s|+6)+|s|(2
m^2+11 )+6). \label{osc}\eea

\section{Additional potential term and related quantum dot models}
 It is obviously seen that one can add to the Hamiltonian (\ref{2.6}) an
 additional potential term of the form
 \begin{eqnarray}
 {\mathcal{U}}=\frac{\omega^2}{2}\left(\rho^2+4 z^2 \right)=\frac{\omega^2}{2}\frac{\xi^3+\eta^3}{\xi+\eta}
 \label{pot_os}
  \end{eqnarray}
 without breaking its classical integrability and separation of
 variables in the quantum case in parabolic coordinates.
 Such kind of potential can originate in the model of a cylindrical quantum dot,
 where the influence of the dot boundary is described by the
 confining potential of the parabolic type (oscillator potential, see also \cite{bmn})
with special rate of
 radial and transverse frequencies. Moreover, if we put $s=0$ the
 emerging system could be identified with the model of a charged
 particle moving in the axially symmetric quantum dot in the field of Coulomb center and
 homogeneous electric and magnetic field pointing along the dot
 symmetry axis.\footnote{
In the noncommutative framework the Coulombic monopole has been
considered in \cite{kh}. }
 The corresponding confining potential takes the
 form
 \begin{eqnarray}
 {\mathcal{U}}_{C}=\frac{1}{2}\left(\omega^2 \rho^2+\tilde{\omega}^2 z^2
 \right), \quad
 \tilde{\omega}=2\sqrt{\omega^2+\omega_B^2}.\label{Uc}
 \end{eqnarray}
This model is integrable at the classical level
\cite{KNO07,NO_TMP} and leads to the separable Schr\"{o}dinger
equation. One can also consider a little different context of that
issue. Let us suppose we have a particle moving in the field of
Coulomb center (or the relative motion of two electrically charged
particles) in the cylindrical quantum dot with an axially
symmetric oscillatory confining potential with arbitrary
frequencies $\omega_{\rho}$ and $\omega_z$. The Schr\"{o}dinger
equation for such a system does not admit separation of variables,
except for the case $\omega_z=4\omega_{\rho}$. Let us assume also
that there are homogeneous electric and magnetic fields both
pointing along the $z$-axis
\begin{eqnarray}
{\mathcal{H}}=\frac{1}{2}\left(p_{\rho}^2+p_z^2+\frac{\left(p_{\varphi}-\frac{1}{2}e
B \rho^2 \right)^2}{\rho^2} \right)+\frac{e
q}{\sqrt{\rho^2+z^2}}-e
{\mathcal{E}}z+\frac{1}{2}\left(\omega_{\rho}^2\rho^2+\omega_z^2
z^2 \right). \label{H_QD}
\end{eqnarray}
However, if the magnetic field magnitude $B$ takes the value
$\frac{1}{e}\sqrt{\omega_z^2-4 \omega_{\rho}^2}$ the variables
separate in parabolic coordinate. So, at the classical level one
obtains an integrable system, which can be a subject for the
perturbation theory in the quantum case. The quantum dots models
with Dirac monopole inside has not only academician interest.
Though, no elementary particle carrying magnetic charge has been
discovered up to now, there are very promising theoretical evidences
of the magnetic monopoles existence as emergent particles, i.e., as
the quasi--particles in some strongly correlated many--body systems.
It was shown in Ref. \cite{SpI} that such kind of magnetic monopoles
do emerge in so--called spin-ice materials, the exotic class of
magnets in which local magnetic moments (spins) are residing on the
sites of pyrochlore lattice (corner shared tetrahedra) and are
constrained to point along their local Ising axes. The spins
interact to each other via nearest neighbor exchange and long--range
dipole--dipole interaction. Thus, the model discussing in this
section can be realized to certain extent in the quantum dots
prepared from spin-ice compound, such as Ho$_2$Ti$_2$O$_7$,
Dy$_2$Ti$_2$O$_7$, e. t. c. (for review of spin-ice see
\cite{SpI_re}).

 In order to apply perturbative results from previous section to
 the quantum dots model discussing here the parameters of
 confining potentials must be small(more precisely, the corresponding correction $\triangle E_{n_1n_2m}$ must be much smaller than
 the distance between corresponding neighbor levels of unperturbed problem), otherwise only the numerical
 calculations are relevant.
  The corresponding correction to the ground state energy has the same form as Eq.
 (\ref{osc}) with $\omega_0=\omega_B+\omega$. In this case, in order for the
 perturbative calculation to be relevant, this
 correction must be much smaller than the ground state energy, which leads to
 \bea
 \omega_0 \ll \left((1+|s|)^3((m^2+s^2)(|s|+6)+|s|(2
m^2+11 )+6) \right)^{-1/2}.
 \eea
  A similar model of the two--electron
 quantum dot subjected to the external homogeneous electric field
 was considered numerically in Ref. \cite{QD}.
\section{Conclusion}
In this paper we considered the quantum mechanical two--center
MICZ--Kepler system in the limit when one of the dyons is situated
at infinity.The electro--magnetic field of such a dyon results in
the homogeneous electric and magnetic fields parallel to each other.
At the same time, the specific additional potential term
(MICZ--term) transforms into oscillator potential in the direction
of the external fields and potential proportional to $\cos \theta$.
Thus, the system considered in the present paper concerns the
"monopolic" generalization of the hydrogen atom (Kepler problem)
subjected simultaneously to the constant uniform electric and
magnetic fields. We separated the variables in the corresponding
Shr\"{o}dinger equation in the parabolic coordinates and analyzed
the simplest case of the emergent MICZ--Kepler--Stark--Zeeman system
when external electric and magnetic fields can be regarded as small
perturbations. The separation of variables is possible in virtue of
the additional potential term. As a starting point we considered the
Hamiltonian of one--center MICZ--Kepler system with well known exact
solution, whereas all other terms were treated as perturbations. The
exact wave functions in parabolic coordinates were used to develop
first order perturbative calculations. We obtained linear in
${\mathcal{E}}$ and $B$ corrections to the unperturbed spectrum of
the ordinary MICZ--Kepler problem. The Stark--effect in the
charge--dyon system was calculated within the perturbation theory up
to second order earlier in series of paper \cite{MNP_04,dipole}.
Here we investigated the first order corrections corresponding to
magnetic field effect (linear Zeeman effect). We found the average
magnetic moment acquired by the charge--dyon system in this case.
For the ground state the induced magnetic moment is even function of
the monopole number $s$,
$\overline{\mathcal{M}}_z(-s)=\overline{\mathcal{M}}_z(s)$ and
vanishes at $s \to \pm \infty$. It is also non--analytic  at $s=0$.
As usually, magnetic field removes the degeneracy of the ground
state . The impact of the additional potential, so--called
MICZ--term, in the first order has much in common with that of the
magnetic field. The sign of the corresponding correction depends
only on the sign of magnetic quantum number $m$. We also analyzed
some condensed matter models which can be relevant in the context of
the quantum mechanical system considered in the papers \cite{KNO07,
QD}. The recent results concerning possibility of formation of the
magnetic monopoles as quasi--particles in some strongly correlated
spin systems, so--called spin--ice materials \cite{SpI} make the
attempts of understanding the behavior of the various
(quantum-)mechanical systems at the monopole background very
important and promising for future experiments. Unfortunately, in
order to obtain reliable results in the quantum dots models with
monopoles in the external fields one can not restrict himself with
integrable cases and/or first order perturbative calculations. One
of the reasons for it is the essential role of the confining
potentials which, generally speaking, can not be regarded as
perturbation. However, the second order corrections in the
MICZ--Kepler--Stark--Zeeman system can lead to new interesting
features, for instance, to the interplay between electric and
magnetic properties of the charge--dyon system. Namely, the
correction to the spectrum proportional to ${\mathcal{E}}B$ which
appears in the second order perturbative calculations means that
average dipole(magnetic) moment depend on magnetic(electric) field.
Another important issue is to understand spin effect and
spin--orbital coupling in the systems with monopoles within the
quantum dots models. We intend to consider all these questions in
the forthcoming papers.

\section{Appendix}
Here we calculate the general integrals which emerge in the
perturbation theory calculations for the quantum mechanical
systems in parabolic coordinates
\begin{eqnarray}
{\mathcal{I}}^{k}_{\pm}=\int_0^{\infty}\int_0^{\infty}\left(\rho_1^k
\pm \rho_2^k\right)f_{n_1, m+s}^2(\rho_1)f_{n_2, m-s}^2(\rho_2)d
\rho_1 d \rho_2
\end{eqnarray}
Let us plug into the integral the expressions for functions
$f_{n,p}$ from Eq. (\ref{fkp}). Then \bea
{\mathcal{I}}^{k}_{\pm}=C^2_{n_1,m+s}C^2_{n_2,m-s}\left(J^k_{n_1,+}J^0_{n_2,-}\pm
J^0_{n_1,+}J^k_{n_2,-} \right), \eea
 where  normalization constant square is
 \bea
C^2_{n,m \pm s}=\frac{(|m \pm s|+ n)!}{n ! (|m \pm s|!)^2}
 \eea
 and
 \bea
 J^k_{n,\pm}=\int_0^{\infty}e^{-\rho}\rho^{|m \pm s|+k}F^2(-n, |m \pm s|+1;
 \rho)d \rho \label{J}
 \eea
 The details of calculations of the more general integral
can be found in \cite{QM}. \bea &&\int_0^{\infty}e^{-k z}z^{\nu-1}
F^2(-n, \gamma; kz )d z= \\ \nonumber &&\frac{\Gamma(\nu)n
!}{k^{\nu}\gamma(\gamma-1)...(\gamma+n-1)}\left(1+\sum_{p=0}^{n-1}\frac{n
(n-1)...(n-p)(\gamma-\nu-p-1)(\gamma-\nu-p)...(\gamma-\nu+p)
}{((p+1)!)^2 \gamma (\gamma+1)...(\gamma+p)} \right)\eea Applying
this formula to Eq. (\ref{J}) one obtains \bea
&&J^0_{n,\pm}=\frac{n
! (|m \pm s|!)^2 }{(|m \pm s|+n)!}=\frac{1}{C^2_{n,m \pm s}} \\
&&J^k_{n,\pm}= \frac{n ! (|m \pm s|+k)! |m \pm s|! }{(|m \pm
s|+n)!}\left(1+\sum^{k-1}_{p=0}\frac{n ! |m \pm s|!
\prod_{q=0}^{2p+1}(k-p+q)}{((p+1)!)^2 (n-p-1)!(|m \pm s|+p+1)!},
 \right), \quad k>0\eea
 Thus
 \bea
 {\mathcal{I}}^{k}_{\pm}=&&\frac{(|m+s|+k)!}{|m+s|!}\left(1+\sum^{k-1}_{p=0}\frac{n_1 ! |m + s|!
\prod_{q=0}^{2p+1}(k-p+q)}{((p+1)!)^2 (n_1-p-1)!(|m +
s|+p+1)!}\right)\pm \\
&&\frac{(|m-s|+k)!}{|m-s|!}\left(1+\sum^{k-1}_{p=0}\frac{n_2 ! |m
- s|! \prod_{q=0}^{2p+1}(k-p+q)}{((p+1)!)^2 (n_2-p-1)!(|m -
s|+p+1)!}\right)
 \eea
 Particularly
 \bea
&&{\mathcal{I}}^{1}_{+}=(|m+s|+1)\left(1+\frac{2 n_1 }{|m+s|+1}
\right)+(|m-s|+1)\left(1+\frac{2 n_2}{|m-s|+1} \right)=2 n \\
&&{\mathcal{I}}^{1}_{-}=|m+s|-|m-s|+2 (n_1-n_2)=2 n_-\\
&&{\mathcal{I}}^{2}_{-}=6 \left(n n_--\frac{m s}{3} \right)\\
&&{\mathcal{I}}^{3}_{+}=(|m+s|+3)(|m+s|+2)(|m+s|+1)+(|m-s|+3)(|m-s|+2)(|m-s|+1)+\\
\nonumber
 &&12\left(n_1 (|m+s|+3)(|m+s|+2)+n_2(|m-s|+3)(|m-s|+2)
\right)+ \\ \nonumber
&&30\left(n_1(n_1-1)(|m-s|+1)+n_2(n_2-1)(|m-s|+1) \right)+\\
\nonumber &&20\left(n_1(n_1-1)(n_1-2)+n_2(n_2-1)(n_2-2) \right)
 \eea
\section{Acknowledgements} We are indebted to Armen Nersessian for
valuable ideas and important comments. We also express our gratitude
to Armen Yeranyan, Hayk Sarkisyan and Pasquale Onorato for useful
discussions and interest toward this work. V. O. thanks the
INFN-Laboratori Nazionali di Frascati where the essential part of
the work was done for warm hospitality. This work was partially
supported by grants NFSAT-CRDF UC-06/07, ANSEF-1386-PS, INTAS under
contract 05-7928, and by the European Community Human Potential
Program under contract MRTN-CT-2004-005104 \textit{``Constituents,
fundamental forces and symmetries of the universe''}.


\begin{thebibliography}{100}

\bibitem{Z} D. Zwanziger, Phys. Rev. \textbf{176}, 1480 (1968).

\bibitem{MIC} H. McIntosh, A. Cisneros, J. Math. Phys. \textbf{11},
896 (1970).

\bibitem{mic_par} L. G. Mardoyan, A. N. Sissakian and V. M.
Ter--Antonyan, Int. J. Mod. Phys. A \textbf{12}, 237 (1997); L. G.
Mardoyan, G. S. Pogosyan, A. N. Sissakian and V. M.
Ter--Antonyan,{\it Quntum Systems with Hidden Symmetry}, MAIK Publ.,
Moscow (2006)

\bibitem{mla} I. M. Mladenov and V. V. Tsanov, J. Phys. A
\textbf{20}, 5865 (1987); J. Phys. A \textbf{32}, 3779 (1999);T. Iwai and Y. Uvano, J. Math. Phys. \textbf{27},
1523 (1986);T. Iwai and Y. Uvano, J. Phys. A \textbf{21}, 4083
(1988); A. Nersessian and V. Ter--Antonyan, Mod. Phys.
Lett. A \textbf{9}, 2431 (1994); Mod. Phys. Lett. A \textbf{10},
2633 (1995).

%\bibitem{iwai} x

%\bibitem{iwai2} x
%\bibitem{terant}x

\bibitem{Lec} A. Nersessian, Lect. Notes Phys. \textbf{698}, 139
(2006).

\bibitem{MNY} L. Mardoyan, A. Nersessian, A. Yeranyan, Phys. Lett.
\textbf{A 366}, 30 (2007).

\bibitem{NO_TMP} A. Nersessian and V. Ohanyan, Theor.
Mat. Fiz. {\bf 155(1)}, 618 (2008).

\bibitem{bny1} S.~Bellucci, A. Nersessian, A. Yeranyan, Phys. Rev. \textbf{D70}, 085013 (2004).

\bibitem{dipol} E. A. Tolkachev, L. M. Tomilchik and Y. M. Shnir,
Yad. Fiz. \textbf{38}, 541 (1983); J. Phys. G \textbf{14}, 1
(1988).

\bibitem{dipole} L. Mardoyan, A. Nersessian, H. Sarkisyan, V.
Yeghikyan, J. Phys. \textbf{A}: Math. Theor. \textbf{40}, 5973
(2007).

\bibitem{MIC1} V. V. Gritsev, Yu. A. Kurochkin, V. S. Otchik, J.
Phys. A: Math. Gen. \textbf{33}, 4903 (2000)

\bibitem{MIC_hyp} A. Nersessian and G. Pogosyan, Phys. Rev. A
\textbf{63}, 020403 (2001).

\bibitem{MIC_any} G. w. V. Meng, J. Math. Phys. \textbf{48},
032105 (2007)

\bibitem{KNO07} S. Krivonos, A. Nersessian, V. Ohanyan, Phys.
Rev. \textbf{D75}, 085002 (2007).

\bibitem{MNP_04} L. Mardoyan, A. Nersessian, M. Petrosyan, Theor.
Math. Phys. \textbf{140}, 958 (2004)

\bibitem{iva} E. Ivanov and O. Lechtenfeld, JHEP \textbf{0309},
073 (2003).

\bibitem{bk} S.~Bellucci, S.~Krivonos, Phys. Rev. \textbf{D74}, 125024 (2006).

\bibitem{BKO07} S. Bellucci, S. Krivonos, V. Ohanyan, Phys.
Rev. \textbf{D76}, 105023 (2007).

\bibitem{new} S.~Bellucci, S.~Krivonos, O. Lechtenfeld, A. Shcherbakov,
arXiv:0710.3832v1 [hep-th], accepted for publication in Phys. Rev.
\textbf{D}, (2008).

\bibitem{old} S. Bellucci, A. Beylin, S. Krivonos, A. Nersessian, E. Orazi
Phys. Lett. \textbf{B616}, 228 (2005); S. Bellucci, A. Nersessian,
Phys. Rev. \textbf{D64}, 021702 (2001); S. Bellucci, A. Nersessian, Nucl. Phys. Proc. Suppl. \textbf{102} 227 (2001).

%\bibitem{abcn} x

%\bibitem{abcn2} x


\bibitem{BKS} S. Bellucci, S. Krivonos, A. Sutulin, Phys.
Rev. \textbf{D76}, 065017 (2007);S. Bellucci, P.-Y. Casteill, A. Nersessian,
Phys. Lett. \textbf{B574}, 121 (2003); S.~Bellucci, A. Nersessian, A. Yeranyan, Phys. Rev. \textbf{D70}, 045006 (2004);
S.~Bellucci, A. Nersessian, A. Yeranyan, Phys. Rev. \textbf{D74}, 065022 (2006); S. Bellucci, S. Krivonos, A. Sutulin, Phys.
Rev. \textbf{D76}, 065017 (2007); S. Bellucci, A. Nersessian,
Phys. Rev. \textbf{D67}, 065013 (2003); Erratum-ibid.
\textbf{D67}, 089901 (2005).


\bibitem{QM} L. Landau, E. Lifschitz, \textit{Quantum Mechanics}, Pergamon
Press (1977)

%\bibitem{bny} x

%\bibitem{bny} x

%\bibitem{bny2} x

%\bibitem{bcn} x

\bibitem{bmn} S.~Bellucci, L. Mardoyan, A. Nersessian, Phys. Lett. \textbf{B636}, 137 (2006).

\bibitem{kh} D. Khetselius, Mod. Phys. Lett. A \textbf{20}, 263
(2005); S.~Bellucci, A. Yeranyan, Phys. Rev. \textbf{D72}, 085010
(2005); S.~Bellucci, A. Yeranyan, Phys. Lett. \textbf{B609},
418 (2005); S.~Bellucci, A. Nersessian, Phys. Lett. \textbf{B542},
295 (2002); S.~Bellucci, A. Nersessian, C. Sochichiu, Phys. Lett. \textbf{B522},
345 (2001).


%\bibitem{by} x
%\bibitem{by2} x

%\bibitem{bn} x

%\bibitem{bns} x

\bibitem{SpI} C. Castelnovo, R. Moessner, and S. L. Sondhi, Nature
\textbf{451}, 42 (2008).

\bibitem{SpI_re} S. T. Bramwell and M. J. P. Gingras, Science
\textbf{294}, 1495 (2001).

\bibitem{QD} N. S. Simenovi\'{c}, R. G. Nazmitdinov, Phys. Rev.
\textbf{B 67}, 041305(R) (2003).





\end{thebibliography}
\end{document}